\documentclass[apjl]{emulateapj}

\usepackage{graphicx}
\usepackage{natbib}
\usepackage{amssymb,txfonts}
\usepackage{multirow}
\usepackage{hyperref}
\usepackage{epstopdf}

\newcommand{\HeII}{\ion{He}{2}}
\newcommand{\Halpha}{H$\alpha$}

\newcommand{\FeIX}{\ion{Fe}{9}}
\newcommand{\FeXII}{\ion{Fe}{12}}

\newcommand{\kms}{km~s$^{-1}$}

\newcommand{\Hinode}{Hinode}
\newcommand{\pixel}{pixel$^{-1}$}

\shorttitle{Prominence dynamics during homologous flares}
\shortauthors{Panesar et al.}

\begin{document}

\title{Destabilization of a Solar Prominence/Filament Field System by a Series of Eight
Homologous Eruptive Flares}

\author{Navdeep K. Panesar\altaffilmark{1}, Alphonse C. Sterling\altaffilmark{2}, Davina E. Innes\altaffilmark{3}, Ronald L. Moore\altaffilmark{1, 2}}

%\author{a\altaffilmark{1,2}, b\altaffilmark{2}, c\altaffilmark{3} and d\altaffilmark{3}}

\affil{{$^1$}Center for Space Plasma and Aeronomic Research (CSPAR), UAH, Huntsville, AL\\ 35805, USA}

%\author{z\altaffilmark{2}}
\affil{{$^2$}Heliophysics and Planetary Science Office, ZP13, Marshall Space Flight Center, Huntsville, AL 35812, USA}
\affil{{$^3$}Max Planck Institut f\"{u}r Sonnensystemforschung, Justus-von-Liebig-Weg 3, 37077 G\"{o}ttingen, Germany}

\email{navdeep.k.panesar@nasa.gov}

%\and

\begin{abstract}
Homologous flares are flares that occur repetitively in the same active region, with similar
structure and morphology.  A series of at least eight homologous flares occurred in active region
NOAA 11237 over 16 - 17 June 2011.  A nearby prominence/filament was rooted in the active region, and 
 situated near the bottom of a coronal cavity.
The active region was on the southeast solar limb as seen from SDO/AIA, and on the disk as viewed
from STEREO/EUVI-B. The dual perspective allows us to study in detail behavior of the prominence/filament
material entrained in the magnetic field of the repeatedly-erupting system.  Each of the eruptions was
mainly confined, but expelled hot material into the prominence/filament cavity system (PFCS).
The field carrying and containing the ejected hot material interacted with the PFCS and caused it to inflate, resulting in a
step-wise rise of the PFCS approximately in step with the homologous eruptions.
 The eighth eruption triggered the PFCS to move outward slowly, accompanied by a weak coronal dimming. 
 As this slow PFCS eruption was underway, a final `ejective' flare occurred in the core of the
active region, resulting in strong dimming in the EUVI-B images and expulsion of a
coronal mass ejection (CME). A plausible scenario is that the
repeated homologous flares could have gradually destabilized the  PFCS, and its subsequent 
eruption removed field above the acitive region and in turn led to the ejective flare, strong dimming, and CME.
\end{abstract}


\section{Introduction}

Prominence eruptions are spectacular solar events. They frequently occur in conjunction with solar
flares and coronal mass ejections (CMEs) \citep{munro79,moore88,low96}. Thus, erupting prominences can be viewed as
indirect indicators of likely-impending CMEs. Observational studies of filament motions can tell us about the evolution
of coronal magnetic field during solar eruptions \citep{shibata95,moore01,sterling12}. It is important to study prominence
eruptions for better understanding of erupting coronal magnetic structures.

Prominences are located along the magnetic neutral line in the highly sheared
core field of enhanced magnetic regions \citep{mackay10,lab10} and
 prominence plasma is plausibly embedded in a canopy
of overlying magnetic field lines, also known as coronal arcades.  Except in the immediate vicinity of the neutral line, the
enveloping coronal arcade is nearly potential (nearly orthogonal to the
neutral line) and confines the prominence plasma and coronal cavity \citep{low94,hudson00}.
A coronal cavity is the feature sometimes visible on the limb as a dark, low-density region
between the prominence and the coronal arcade \citep{gib06,gib10}. The coronal cavity is most visible when the prominence and coronal arcade are viewed end-on.
 Sometimes the cavity is observed on top of the cool prominence material \citep{regnier11,panesar14c} and sometimes
 surrounding the prominence \citep{berger12,panesar13,schmit13}.

 \begin{figure*}
   \centering
    \includegraphics[width=0.95\linewidth]{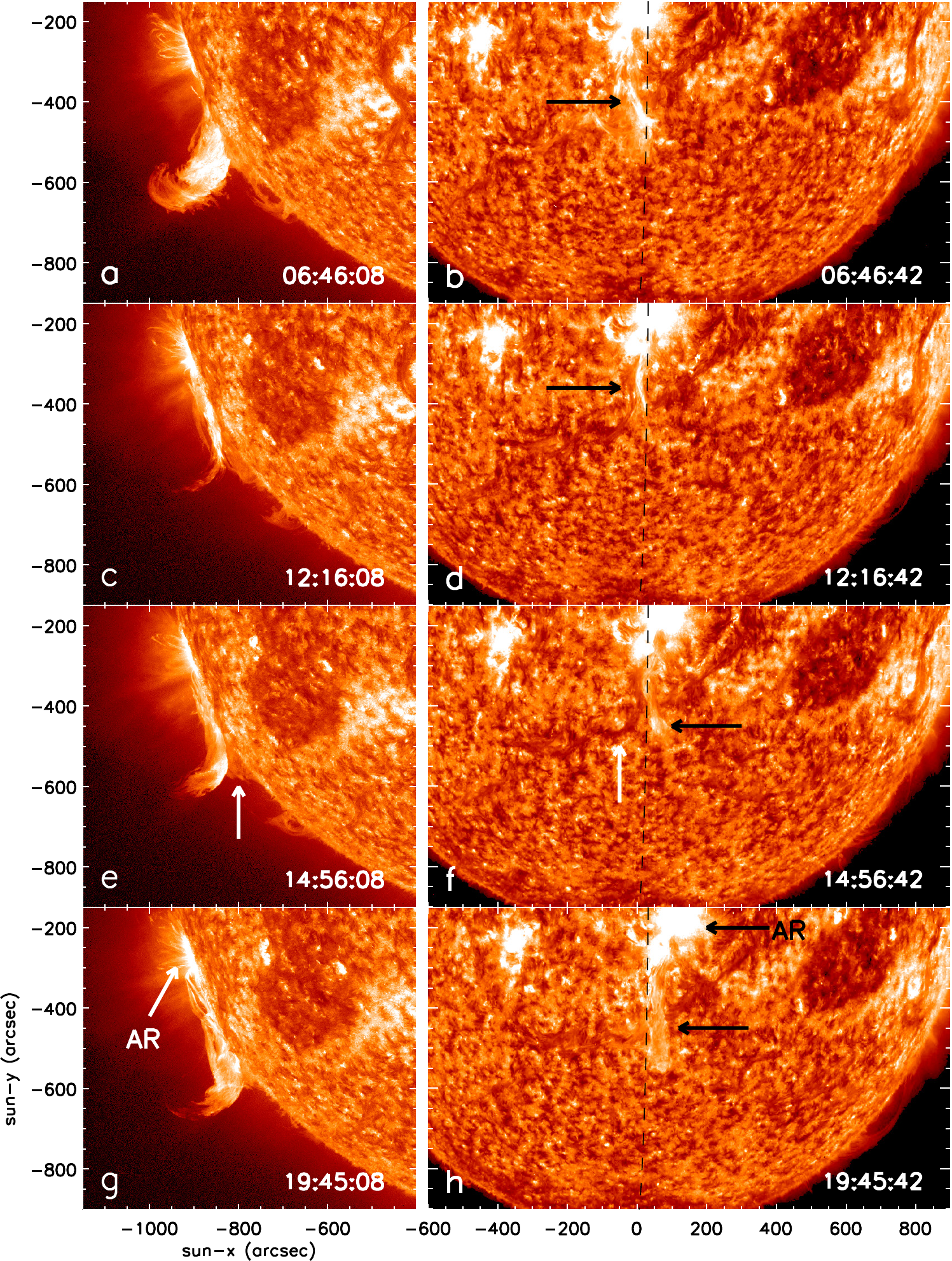}
     \caption{Homologous flares observed on 16 June 2011: (left) AIA 304 \AA\ intensity images; (right) corresponding
     EUVI-B 304 \AA\ intensity images.
  On the EUVI images, the black dashed line shows the position of the AIA limb in the images on the left. In (e) and (f), the white arrows point
  to the prominence/filament cavity system (PFCS). In (b), (d), (f) and (h),
  the black arrows show hot material ejected from the core of the active region
  during the flares. The Y-axis is the same for all images in both columns.
  North is upward and west is to the right in these images, and
  in all other solar images in this paper.} \label{over}
\end{figure*}

\begin{figure*}
   \centering
  \includegraphics[width=0.95\linewidth]{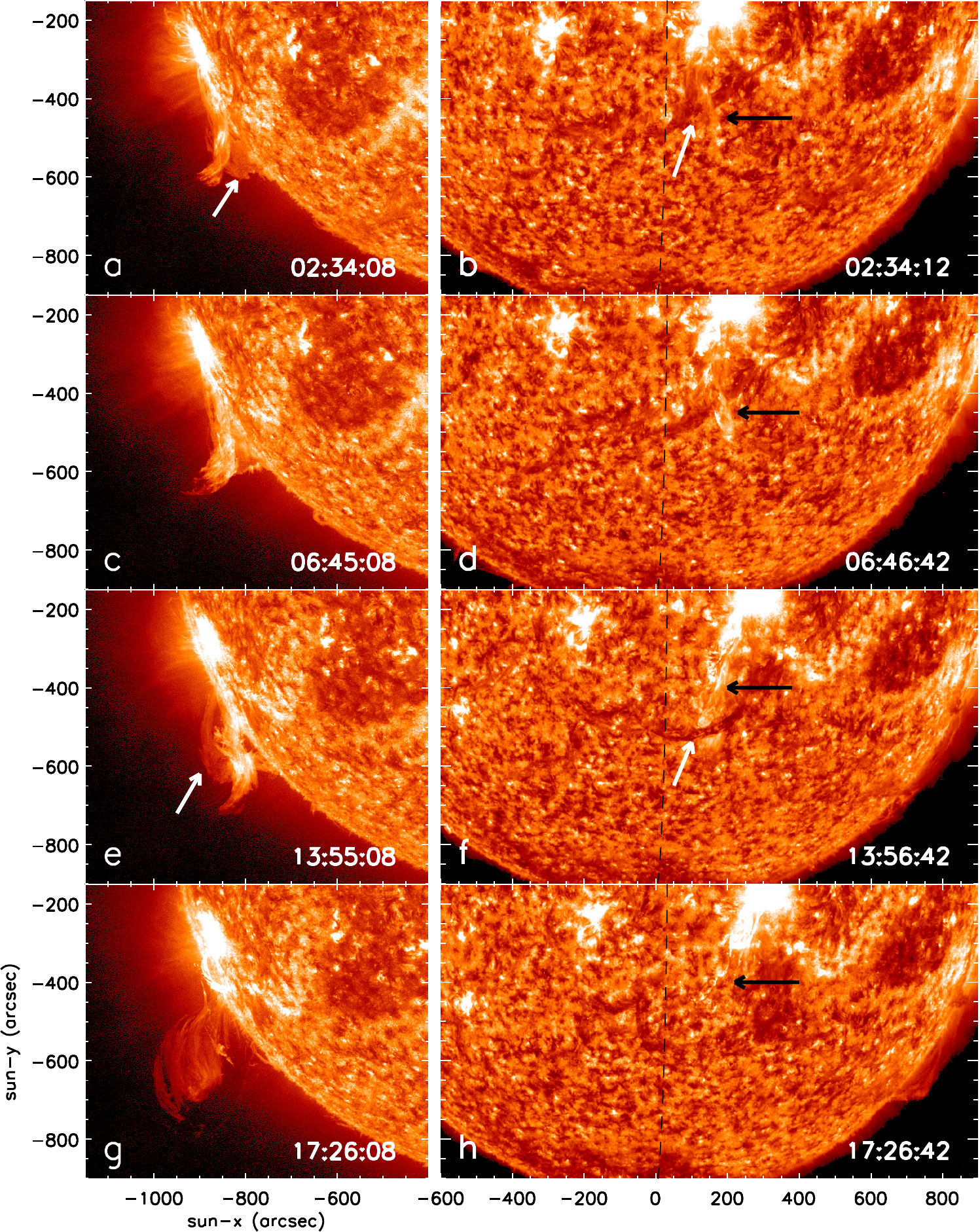}
     \caption{Homologous flares observed on 17 June 2011: (left) AIA 304 \AA\ intensity images; (right) corresponding EUVI-B 304 \AA\ images.
   In (e) and (f), the white arrows point to the erupting prominence/filament.
   Annotated features are same as those in Figure~\ref{over}.} \label{over1}
\end{figure*}

\begin{figure*}
   \centering
  \includegraphics[width=\linewidth]{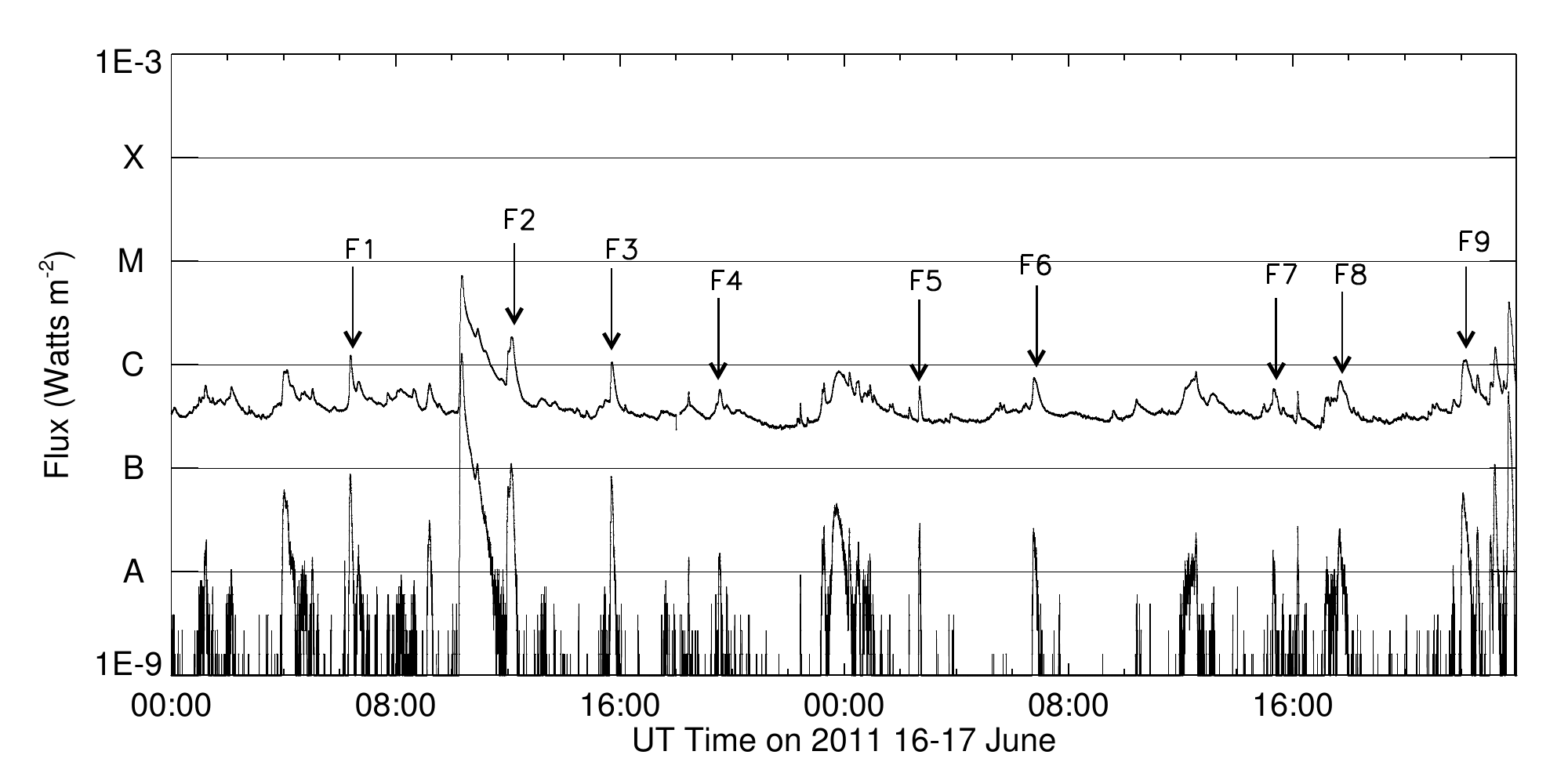}
     \caption{\textit{GOES} soft X-ray flux in 1.0 - 8.0 \AA\ and 0.5 - 4.0 \AA\ passbands from 16 to 17 June 2011
     showing the nine flares discussed in this paper from
     active region NOAA 11237. The arrows are drawn at the flare (F1 to F9) times.} \label{flux}
\end{figure*}

Highly sheared magnetic field plays an
important role in the equilibrium of prominences.
Processes such as magnetic reconnection cause the low-lying sheared field to erupt, carrying the prominence material with it \citep{Balle89,Balle00}.
A prominence is in equilibrium when the downward magnetic tension force balances the upward magnetic pressure force \citep{Priest89,low95}.
If this force balance becomes disrupted, the field may lose its equilibrium and give rise to the initiation of an eruption. However, it is
not necessary that all of the prominence material escapes the Sun.
Observational studies show a wide range of
prominence eruptions, e.g., ejective and
confined eruptions \citep{ji03,sterling12,kuridze13}; see also \cite{torok05} for MHD simulations of coronal flux ropes.
In the case of confined eruptions a filament is seen to rise and abortively erupt without a CME, whereas
in ejective eruptions a filament fully erupts along with a CME \citep{gil01,moore01,torok05}.
It has been reported by \cite{munro79} that 70\%
of CMEs involve prominence eruptions or disappearances.
A large number of theoretical models have been proposed to explain the
initiation of solar eruptions, e.g., tether cutting \citep{moore01}, magnetic breakout \citep{Antiochos99} and flux emergence \citep{chen00},
etc. Based on observations, it has been suggested that ejective eruptions produce long-duration flares
 in comparison to confined eruptions \citep{wang07,cheng11}.

Coronal dimmings are indirect signatures of CME expulsions in coronal images. It has been reported that dimmings can begin
$\sim$ 30 minutes before the appearance of CMEs \citep{stewart74a,stewart74b,rust76,harrison00}. In some cases they begin $\sim$ 3 hours before the
CME onset \citep[e.g.][] {gopal01}.
In addition to accompanying CMEs, they also accompany filament and prominence eruptions \citep{gopal98,harrison00,kiepenheuer64,harrison03}. 
Basically, coronal dimming is a drop in emission due to the
 removal/transport of coronal plasma during eruptions. It has been observed that the dimmings can cover either small or
  large areas of the Sun \citep{zhukov07,Innes10}.

 Prominence eruptions have rarely been discussed in the context of homologous
flares \citep{chandra11,bocchialini12}. These are flares that occur repetitively from the same active region, with similar shape,
morphology, and footpoints \citep{waldmeier38,woodgate84} as seen in  \Halpha ,
EUV and in soft X-ray emission \citep{martres84,martres89,ranns00,sterling01,chandra11}.
It is an important phenomena to study because it sheds light on the conditions of the flare trigger processes and on the mechanisms of
energy storage and release. These types of series of events suggest that either there is a continuous amount of magnetic energy
supplied to an active region, or that not all of the available free energy is released during the eruptions \citep{schrijver09}.
During homologous flares continuous flux emergence in the flaring active
region has been reported by \cite{ranns00}, \cite{sterling01} and \cite{nitta01}.

In the present work, we report on observations of eight homologous eruptive flares that took place adjacent to a prominence.
We mainly focus on the dynamics of the prominence/filament
cavity system driven by the flare eruptions using the observations recorded by
the Atmospheric Imaging Assembly (AIA) onboard the Solar Dynamics Observatory (SDO), and  the
Extreme UltraViolet Imager (EUVI) onboard the Solar TErrestrial RElations Observatory (STEREO).
We will show that each of the eight flares was a confined eruption, with each
eruption expelling field into the prominence/filament field system. The expelled field
 interacted with the prominence/filament cavity system, and each of 
the homologous eruptions thereby successively relaxed the stress holding down that
prominence/filament cavity. Consequently, the prominence/filament cavity rose in height in
successive steps, until it became destabilized and erupted after the final
homologous event.  This eruption of the prominence/filament cavity system apparently removed overlying field from the
homologous-flare-producing active region, so that the next eruption from that active
region was not confined. We infer that the prominence/filament cavity eruption in combination with the final
eruption from the active region resulted in a CME that we observed in
SOHO/LASCO. To the best of our knowledge, prominence eruptions triggered by homologous flares have never been
reported before.

\begin{figure}
   \centering
  \includegraphics[width=\linewidth]{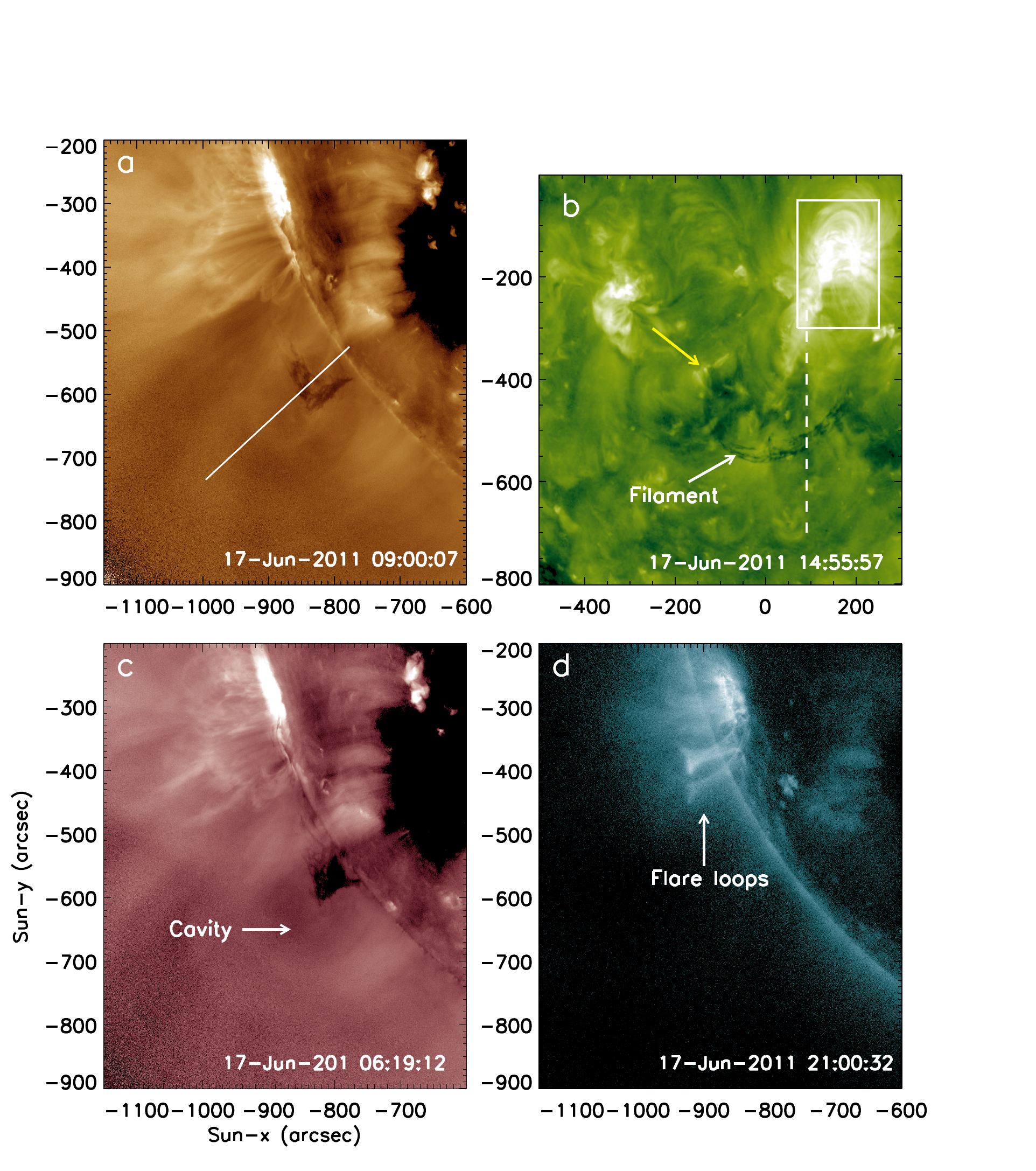}
     \caption{Prominence/Filament cavity system: a), c) and d) show AIA 193, 211 and 94 \AA\ intensity images, respectively; b)
     shows EUVI-B an 195 \AA\ intensity
     image. In (a), the white diagonal line shows the position of the time-series in Figure~\ref{mosac}a. In (b),
     the white dashed line shows the position of the time-series in Figure~\ref{mosac}b and \ref{dim}.
     The yellow arrow in (b) indicates the
     brightening at the remote footpoint of the filament during F7. The boxed region in (b) shows the
     location of the light curve in Figure~\ref{mosac}c.
     } \label{4sdo}
\end{figure}

\section{Instrumentation and data}\label{data}

The eight homologous flares were observed by SDO/AIA \citep{lem12} on the southeast solar limb. Each flare was concurrently
observed  as an on-disk event occurring near the central meridian in STEREO/EUVI-B \citep{Howard08} images.
They occurred in active region NOAA 11237 over 16 - 18
June 2011. We study the dynamics of a prominence/filament cavity system that was magnetically connected to the
 flare eruptions by combining the
observations from the two directions. During this period (16-18 June 2011),
\Hinode\ was observing elsewhere, and therefore has no data for
these events.

Our main data for the limb view are EUV images from SDO/AIA.
AIA provides full-Sun images with a high spatial resolution (0.6\arcsec\ \pixel) and high temporal cadence ($\sim$ 12s)
in seven EUV wavelength bands \citep{lem12}. We used the images from three channels: 304 \AA, which is centered
around an \HeII\ line (0.05 MK); 171 \AA, which is dominated by an \FeIX\ line (0.7 MK); and 193 \AA,
which is centered at an \FeXII\ line formed around 1.5 MK, but also has some response to 0.25 MK plasmas \citep{del11}.
For the analysis we mainly used 1- and 5-minute cadence image sequences.
We removed the average coronal background from 171 and 193 \AA\ images to enhance the visibility of the coronal cavity and overlying
coronal structures. The background image was computed by taking the mean of three months of data over May - July 2011.
 We also examined in a cursory fashion images from the 211 \AA\ and 94 \AA\ channels,
formed around 1.9 MK and 6.0 MK, respectively.

Magnetograms from the Helioseismic and Magnetic Imager (HMI) aboard SDO
 are also used to examine the photospheric magnetic field. HMI provides full-disk line-of-sight magnetograms
with high spatial resolution of 0.5\arcsec\ pixel$^{-1}$ and temporal cadence of $\sim$ 45s \citep{scherrer12}.
A few HMI magnetograms
have been used to examine the line-of-sight photospheric magnetic field of the region of interest on 19 June 2011, when the active region was
clearly visible on the disk.

For the disk prespective, we used 195 \AA\ and 304 \AA\ EUVI-B images with a spatial resolution of 1.6\arcsec\ ~\pixel~ and temporal cadence of $\sim$ 5-minutes
and $\sim$ 10-minutes, respectively.
The STEREO 195 \AA\ and 304 \AA\ images are centered around the same \FeXII\ (1.5 MK) and \HeII\ (0.05 MK) lines
as the AIA 193 \AA\ and 304 \AA\ images. 
To identify mutual features in the EUVI-B on-disk-perspective images and in the
AIA limb-perspective images, we first selected sub-images of about the same-sized
field of view  from both instruments, centered on approximately the same
feature as seen from the two perspectives.  Since EUVI-B
has much courser cadence  than AIA, we then selected out AIA images that were
from about the same time as the corresponding EUVI-B images.  With these images we
constructed movies (of $\sim 5$~min cadence) from the two instruments.  Viewing
the so-processed pair of movies (especially but not exclusively in corresponding
wavelengths, meaning 304~\AA\ images for AIA and EUVI-B, and 193~\AA\ and
195~\AA\
respectively for AIA and EUVI-B) greatly facilitated identification of the mutual
features. The STEREO images have been derotated to a particular time.
%In Figures~\ref{over} and \ref{over1}, the SDO limb line is plotted on the STEREO images to  show the exact position of the AIA limb features on the STEREO disk images.
To show the exact position of the AIA limb features on the EUVI-B 304~\AA\ disk images,
we plotted the SDO limb line on the STEREO images (Figure~\ref{over} and Figure~\ref{over1}).
We also created EUVI-B
base-difference images to show better changes in faint structures. We created movies from the various sets of images, to
study the events in detail.

\begin{figure}
   \centering
  \includegraphics[width=\linewidth]{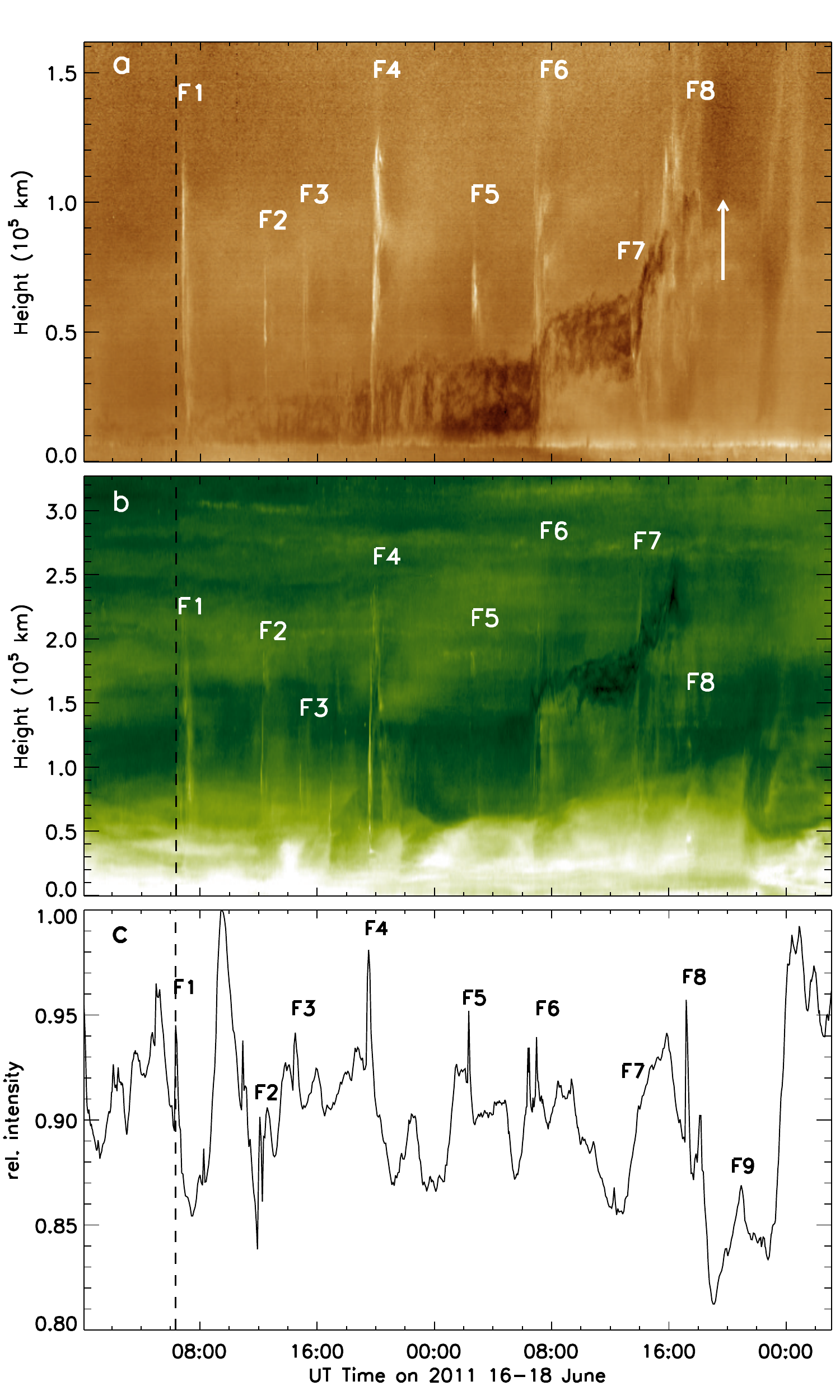}
     \caption{Prominence/Filament trajectories: a) AIA 193 \AA\ intensity time-series image along the diagonal line in Figure~\ref{4sdo}a;
     b) EUVI-B 195 \AA\ intensity time-series image along the dashed line in Figure~\ref{4sdo}b; c) EUVI-B 195 \AA\ integrated intensity
     plot of the region selected in the box shown in Figure~\ref{4sdo}b. Nine flares are labeled as F1-F9.
     In (a), the arrow points to the weak dimming after F8.} \label{mosac}
\end{figure}

\begin{figure}
   \centering
  \includegraphics[width=\linewidth]{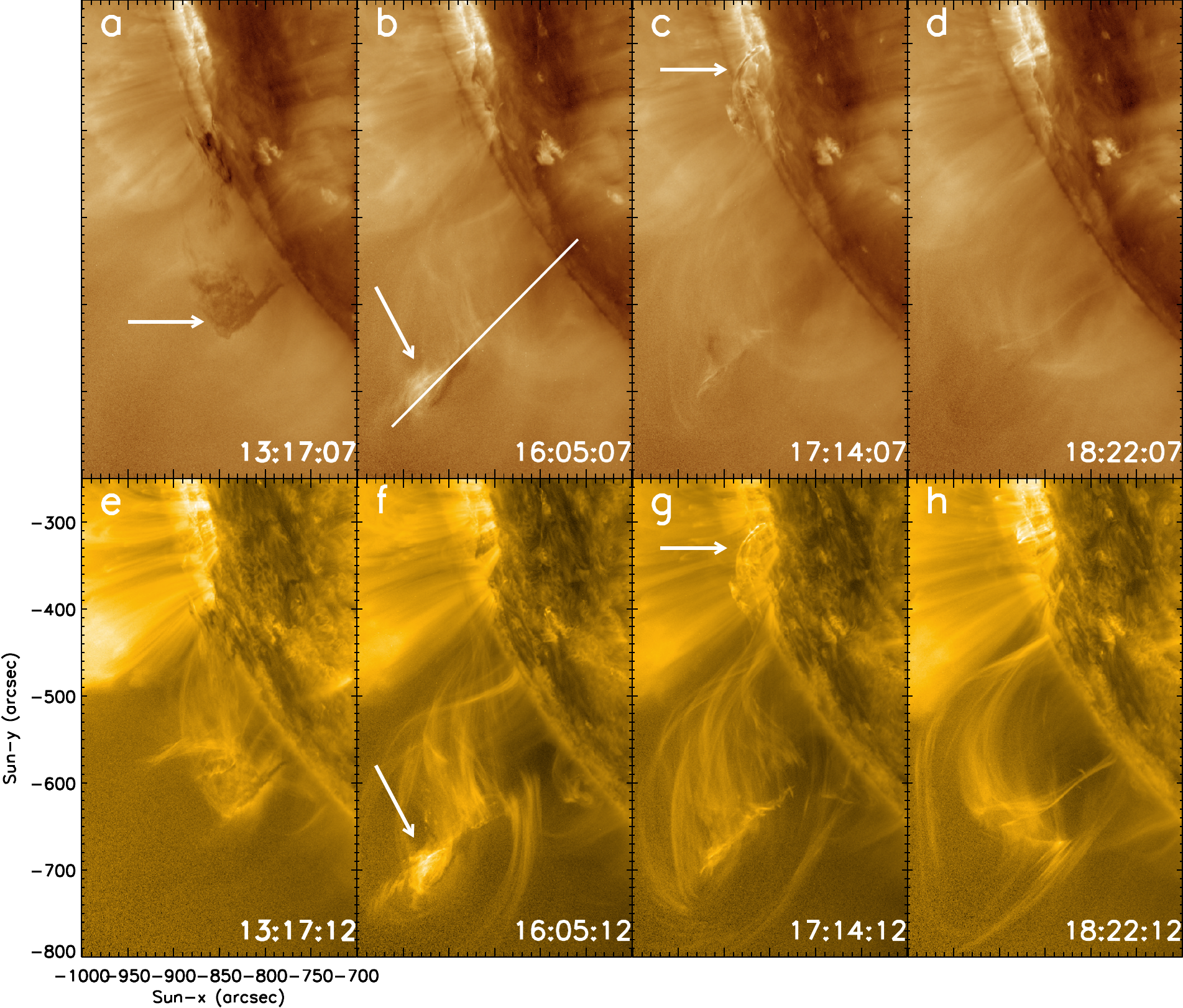}
     \caption{Evolution of prominence on 17 June 2011: (a-d) AIA 193 \AA\ intensity images; (e-h) AIA 171\AA\ intensity images.
  The arrow points to the prominence in (a). In (b) and (f), the arrows point to the
   brightening at the top of the prominence.
  The diagonal line in (b) marks the position of the time-series shown in Figure~\ref{reconnection1}a,b.
  The arrows in (c) and (g), indicate the brightening in the core of the active region during F8.} \label{reconnection}
\end{figure}

\section{Observations}\label{observation}

\subsection{\textit{The eight homologous flares}} \label{homo}

In Figures~\ref{over} and \ref{over1}, we show AIA 304 \AA\ and EUVI-B 304 \AA\ images of the eight homologous flares from
active region NOAA 11237 on 16 and 17 June 2011. Over that period, NOAA 11237 was a matured active region on the eastern limb of the Sun.
Figure~\ref{flux} shows the \textit{GOES}-15 soft X-ray flux time profile from 16 to 17 June 2011. From this we can obtain information about the
X-ray classes of the flares and the peak-intensity times.
We identified nine peaks, during nine flares from
 NOAA 11237, which we identified from the movies (MOVIE304 and MOVIE193) and flare lists\footnote{$http://hesperia.gsfc.nasa.gov/hessidata/dbase/hessi$\_$flare$\_$list.txt$}$^{,}$ \footnote{$http://www.lmsal.com/solarsoft/latest$\_$events$\_$archive.html$}.
 The nine flares are highlighted with the arrows from F1 to F9.
 The first eight of these (F1-F8) correspond to the eight homologous flares shown in Figures~\ref{over} and \ref{over1}, and the ninth (F9)
 corresponds to the ejective flare from the region.
  Four of these flares were on 16 June 2011: the first three were C-class and
the last one was of B-class.
  The remaining five were on 17 June 2011: the first four were B-class and the fifth was C-class.
  Besides F1-F9, a few other small peaks in the GOES
plot (including the C flare just before F2) were from the same active region, but appear to be independent of the features we are focusing on here;
some of those brightening are discussed near the end of Section~\ref{homo}
Apart from the studied active region, three more active regions were
visible during these days. Some other significant peaks in the GOES plot are due to activity from the other three active regions.

We observed a prominence/filament cavity system (PFCS) adjacent to the active region, shown with white
 arrows in Figure~\ref{over1}a,b. The cavity was discernible only for a few hours on 17 June 2011, surrounding the prominence. It is
 apparent in the 193\AA\ and 211\AA\ images (Figure~\ref{4sdo}a,c).
 When referring to the cool material of the PFCS in this paper, we use the term `prominence' for the limb view,
 `filament' for the disk view, and `prominence/filament'
 for both limb/disk views.
 The core of the active
 region was separated by $\sim$ 200\arcsec\ from the prominence's spine. 
 From STEREO images, we can see that the prominence is behind the limb from the SDO perspective on 16 June 2011 (Figure~\ref{over}b).
 One of the footpoints of the prominence/filament is anchored in the active region (Figure~\ref{over}d) and
 the other footpoint is southeast of the active region in a quiet-Sun location, as we can see in the STEREO images (Figure~\ref{over1}b).
 The active region and prominence/filament is the part of the same system. 
 To relate the line-of-sight magnetograms
with on-disk structures in the EUVI-B images, we looked at HMI images taken two days later,
on 19 June 2011, when the active region and magnetic neutral
line were more clearly visible on the SDO disk images.
For this purpose, we summed 30-minute’s worth of 45-second cadence co-aligned HMI magnetograms, thereby building a single deep magnetogram.
The filament and its far footpoint (i.e the footpoint farthest from the active region, shown in Figure~\ref{4sdo}b) resided in a weak-field region,
based on the SDO/HMI line-of-sight
 magnetogram of the region on 19 June 2011 (Figure~\ref{mag}).

\begin{figure}
   \centering
  \includegraphics[width=\linewidth]{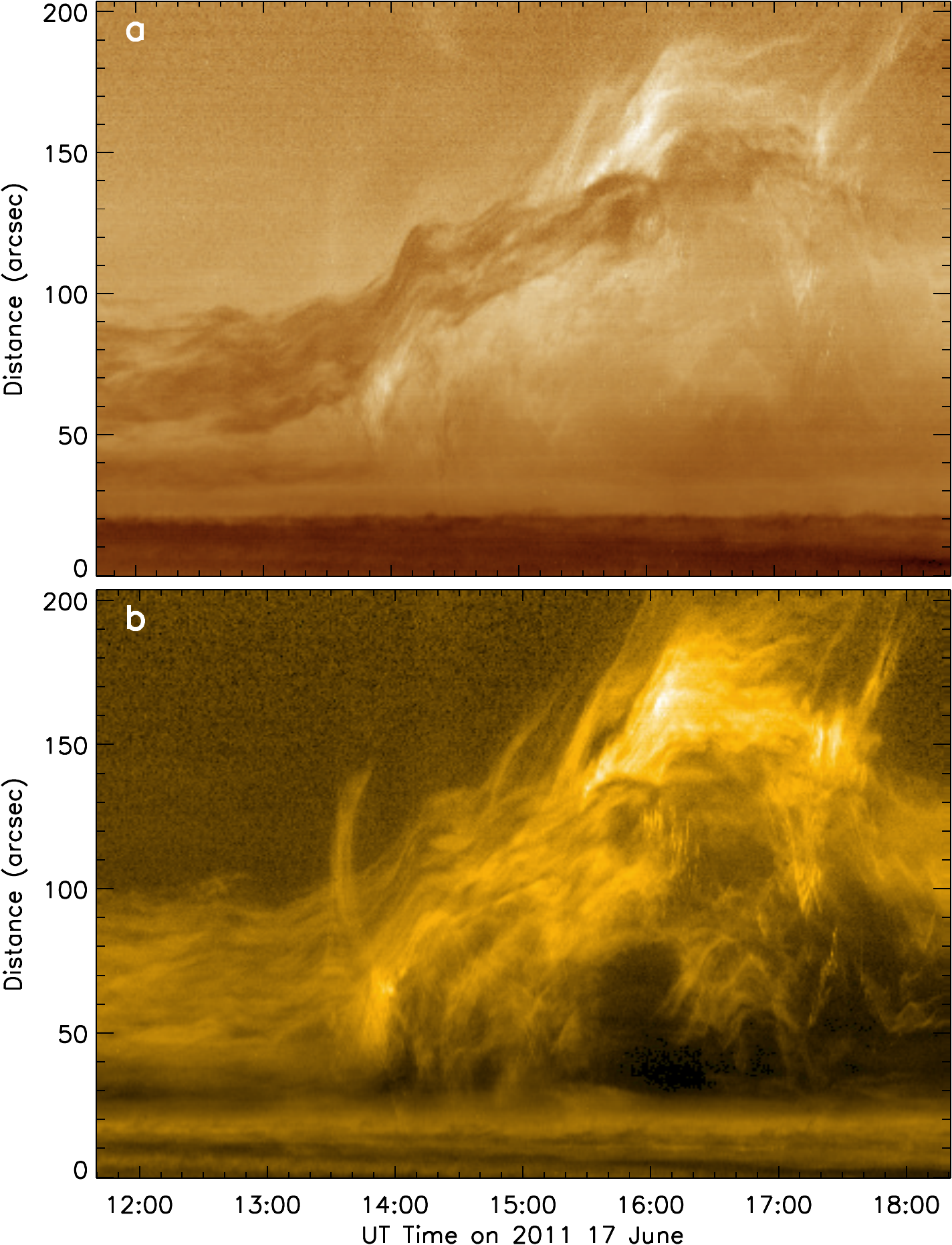}
     \caption{Prominence eruption on 17 June 2011: (a) AIA 193 \AA\ and (b) 171 \AA\ intensity time-series
     image along the diagonal line in
     Figure~\ref{reconnection}b. Initially, the prominence rises at 13:50 UT, and
     then it starts to fall toward the solar surface at 17:05 UT.} \label{reconnection1}
\end{figure}

 Here we discuss in more detail the nine flares from the active region and the material ejected along with
 these flares, as seen from AIA and EUVI-B.
  Figure~\ref{over} and \ref{over1} show the concurrent brightenings in both images during the flares
 (shown with black arrows on STEREO images). The repeated flares from the same active region, from the same position, certainly come under the
 category of homologous flares, discussed by, e.g.\cite{woodgate84} and \cite{choe00}. A close-up view of the homologous flares
 is shown in Figure~\ref{flares}a,b and in MOVIEAIA. 
 All of the eight homologous flares were accompanied by confined eruptions.
 Each of the homologous flares had the same pattern, with a brightening at the flaring location,
 followed by ejection of material (shown with black arrows in Figure~\ref{over}b and \ref{over1}b) and field
 from the flaring region. That ejecta interacted with the PFCS, and
 successively disrupted the overall stability of that system.
 In Figures~\ref{over1}e and ~\ref{over1}f, the white arrows point to the activated
 prominence/filament after flare F8. The flares' onsets and related prominence/filament response can be seen in
 the 304\AA\ movie accompanying Figure~\ref{over} and \ref{over1}.

 A note on our use of ``confined"- and ``ejective"-eruption terminology:  In this paper we say that
the eight homologous flares, F1-F8, are "confined," even though each of these eruptions expelled
ejecta. More specifically, our meaning is that each of these eight flares ejected material only out of their immediate vicinity,
 i.e. the ejecta left the active region in which those flares occurred.  But,
in each of these eight cases, the so-ejected material became trapped within the field of the
PFCS, and therefore that material did not leave the Sun and form a CME, as in
canonical ejective flares.  In this sense, the eight homologous flares were not completely confined or
completely ejective, as used in standard solar terminology.  Our convention here is to say that
they were confined, because their ejected material did not escape from the Sun as a CME.  This
is in contrast to the case of flare F9, which was clearly ejective because its expelled material left
the Sun, as we will discuss later.

To explain the general progression of the homologous eruptions,
 in this paragraph we discuss the progression of events around the time of homologous flare F6;
 we refer the reader to MOVIE193 accompanying this paper. The flare started at 06:20 UT on 17 June 2011.
 There was a brightening in the south of the active region
 (at the magnetic neutral line, Figure~\ref{mag}) at 06:20 UT, see MOVIE193.
 At 06:25 UT some hot material, together with magnetic field, was ejected
from the core of the active region as typically happens according to the standard model for solar eruptions 
\citep[e.g.][]{shibata95,moore01}.
 %A cool prominence material  ejected from the core followed by the bright ejecta which is clearly visible in AIA 193\AA\ images (MOVIE193 at 06:26 UT).
 At the same time the ejection of the hot material is also seen in the EUVI-B 195 \AA\ images as bright ejecta.
 This bright ejecta hit the
 prominence/filament at 06:45 UT and resulted in an upward rise of the prominence/filament. The ejecta became trapped in the cavity
 field lines, which were connected to the PFCS. The AIA limb images clearly show the movement of
 bright material along the cavity field lines at 06:50 UT. Simultaneously,
 we can see that the brightening swept over
 the filament and the remote footpoint of the filament brightened (indicated by an arrow in Figure~\ref{4sdo}b during F7).
 The brightening is faintly visible in the EUVI-B frames of
 MOVIE193 at 07:25 UT, which is after F6. 
 All eight homologous events followed this same basic pattern with only slight variations.
 In particular, after F8, the prominence/filament disappeared and the cool material drained
 to the solar surface. The field containing the PFCS continued to  expand slowly after F8 and completely removed
 overlying field from above the active region.

\begin{figure}
   \centering
  \includegraphics[width=\linewidth]{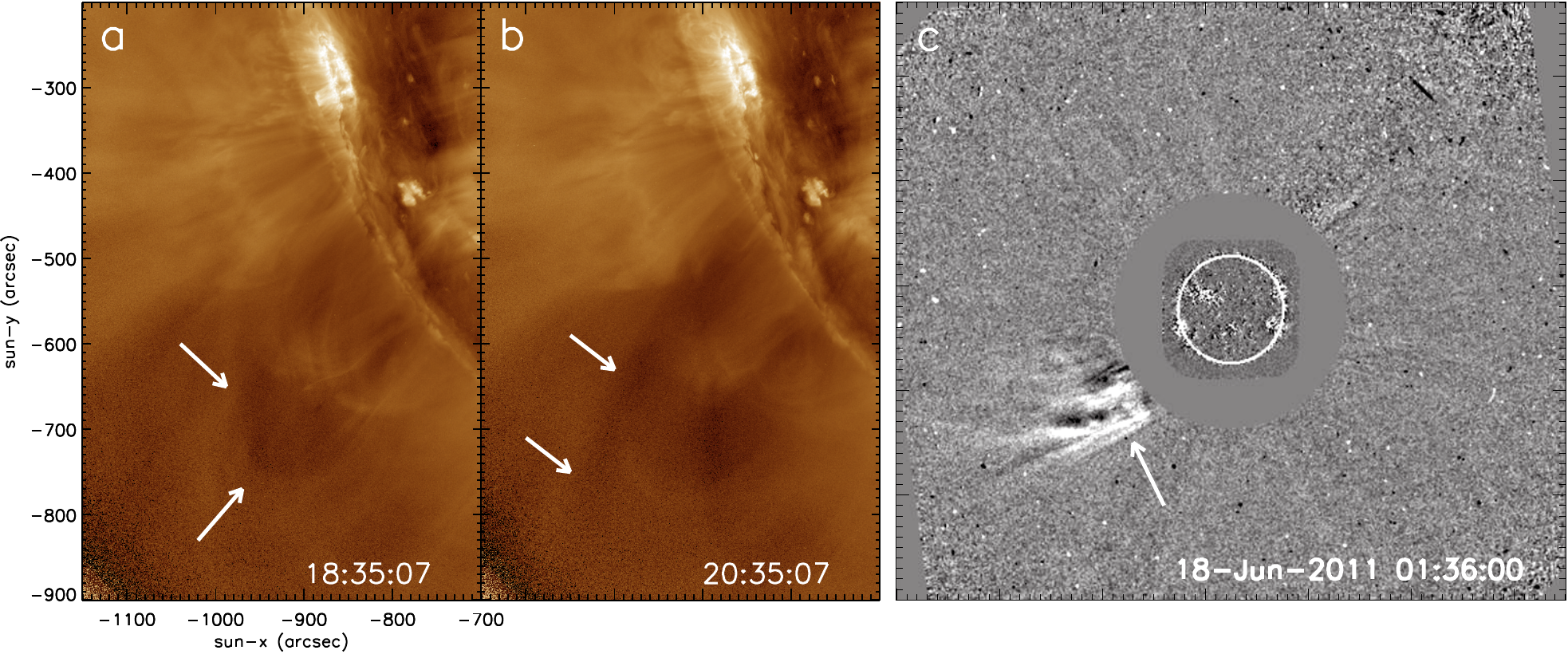}
     \caption{Coronal arcade and CME: (a-b) AIA 193 \AA\ intensity images. The white arrows point to the
     expansion of the overlying coronal field after F8. (c) LASCO (C2) and AIA 193\AA\ running-difference
     image; an arrow indicates the CME after F9.} \label{arc}
\end{figure}

 The last flare, F9, occurred at $\sim$ 21:00 UT. It was an ejective eruption (flare loops are shown in Figure~\ref{4sdo}d).
 A CME followed the last flare, F9.  The CME is detected by LASCO as shown in Figure~\ref{arc}c.
 According to the LASCO catalog\footnote{$http://cdaw.gsfc.nasa.gov/CME$\_$list$}, the CME material had a velocity of $\sim$ 234 \kms.
 If we follow the linear fit of the plane-of-sky velocity back in time, it
 matches well with the flare (F9) timing. This suggests that the CME occurred in conjunction with F9.

 To see the relation between prominence/filament motion and the flares, we construct height-time plots over
 two cuts: one along the prominence in AIA images (Figure~\ref{4sdo}a), and the other across the filament in EUVI-B images (Figure~\ref{4sdo}b).
 Figure~\ref{mosac}a,b shows the height of the filament as a function of time, along the fiducial lines plotted in Figure~\ref{4sdo}a,b.
 In the time-series image, the eight homologous flares are marked as F1 to F8. The filament trajectory is not visible before F6 in
 Figure~\ref{mosac}b, because the position of the fiducial line is close to the active region, away from the center of the filament.
 We selected the location for the EUVI-B fiducial in Figure~\ref{4sdo}b so that the bright ejecta from the flares in the
 active region would cross that fiducial; these bright ejecta are labeled as F1-F8 in Figure~\ref{mosac}a.
 In other words, it is the bright ejecta from the flares
 crossing the fiducial line that appear as
 F1-F8 in Figure~\ref{mosac}a,b, but the filament itself does not cross the fiducial line (Figure~\ref{4sdo}b) as it
 moves prior to F6. Those early-time
 prominence motions are however well captured in the AIA height-time plot (Figure~\ref{mosac}a), which are based on the fiducial line
 in Figure~\ref{4sdo}a.

  In Figure~\ref{mosac}a,b, we show the flare brightenings (vertical bright lines labeled F1-F8)
  during the eight homologous eruptions, interacting with the prominence/filament. Figure~\ref{mosac}a,b shows the relation between the flares and
  prominence/filament. 
  There is a slow rising motion of the prominence/filament apparent in both SDO and STEREO time-series images (Figure~\ref{mosac}a,b).
 There are conspicuous jumps in the prominence trajectory after F4, F6 and F7. The prominence does not show any
 increase in height during F3 and F5 (Figure~\ref{mosac}a).
 Before F3, the prominence was beyond the SDO limb, therefore it is not
 clearly visible in SDO time-series image (in Figure~\ref{mosac}a).
 Figure~\ref{mosac}a covers three important dynamics of the prominence: (1) the prominence enters its 
 fast-rise phase at 13:50 UT on 17 June 2011, (2) the prominence disappears
 completely with F8, and (3) a coronal dimming (indicated with an arrow in Figure~\ref{mosac}a) starts after the eruption F8.

\begin{figure}
   \centering
  \includegraphics[width=\linewidth]{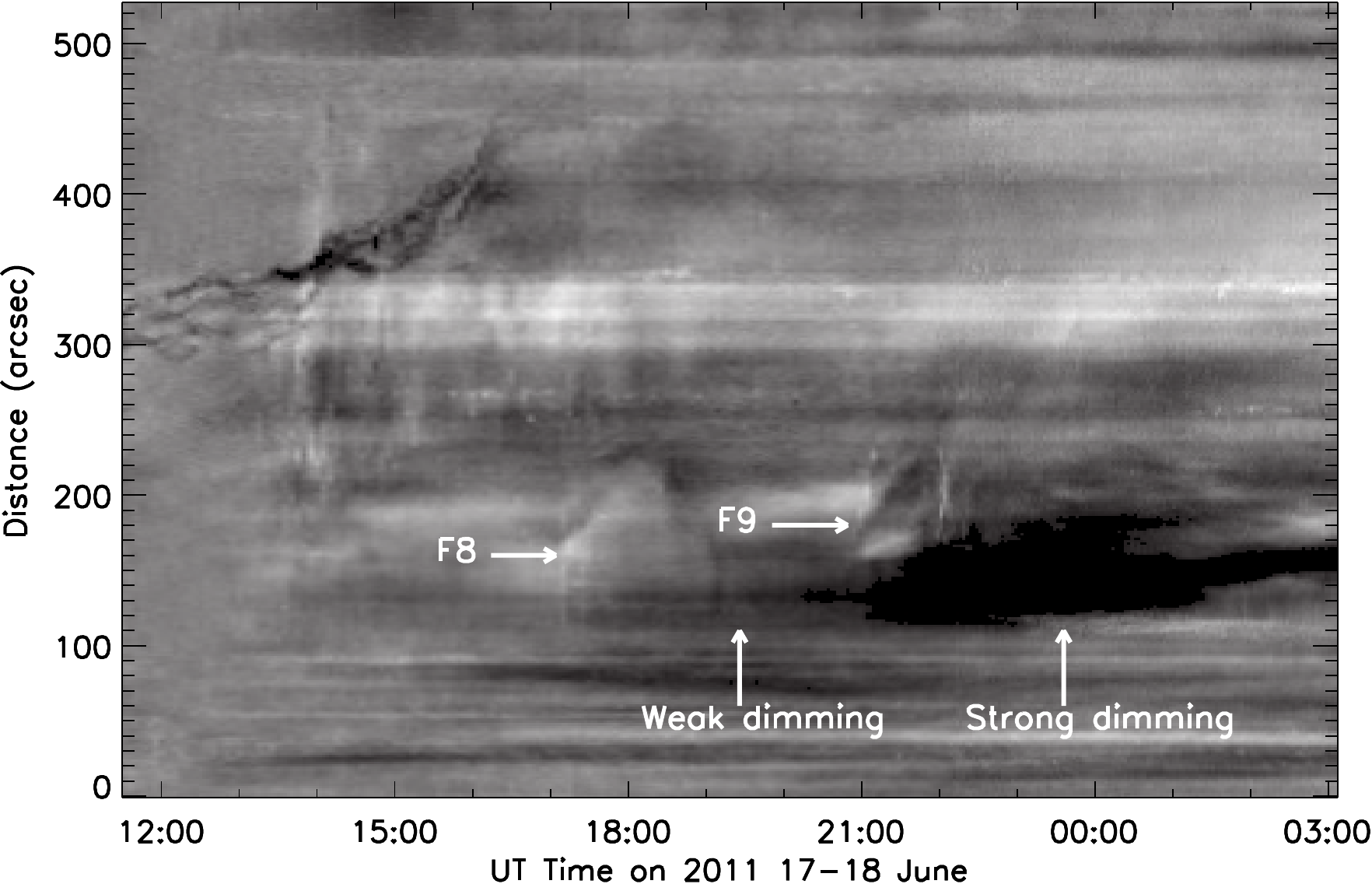}
     \caption{Coronal dimming: EUVI-B 195 \AA\ base-difference time-series image along the dashed line in Figure~\ref{4sdo}b.
      The weak and strong coronal dimming is apparent after F8 and F9, respectively.
      The subtracted base image time was obtained by averaging images between 11:45 and 13:00 UT.} \label{dim}
\end{figure}

\begin{figure}[h]
   \centering
  \includegraphics[width=\linewidth]{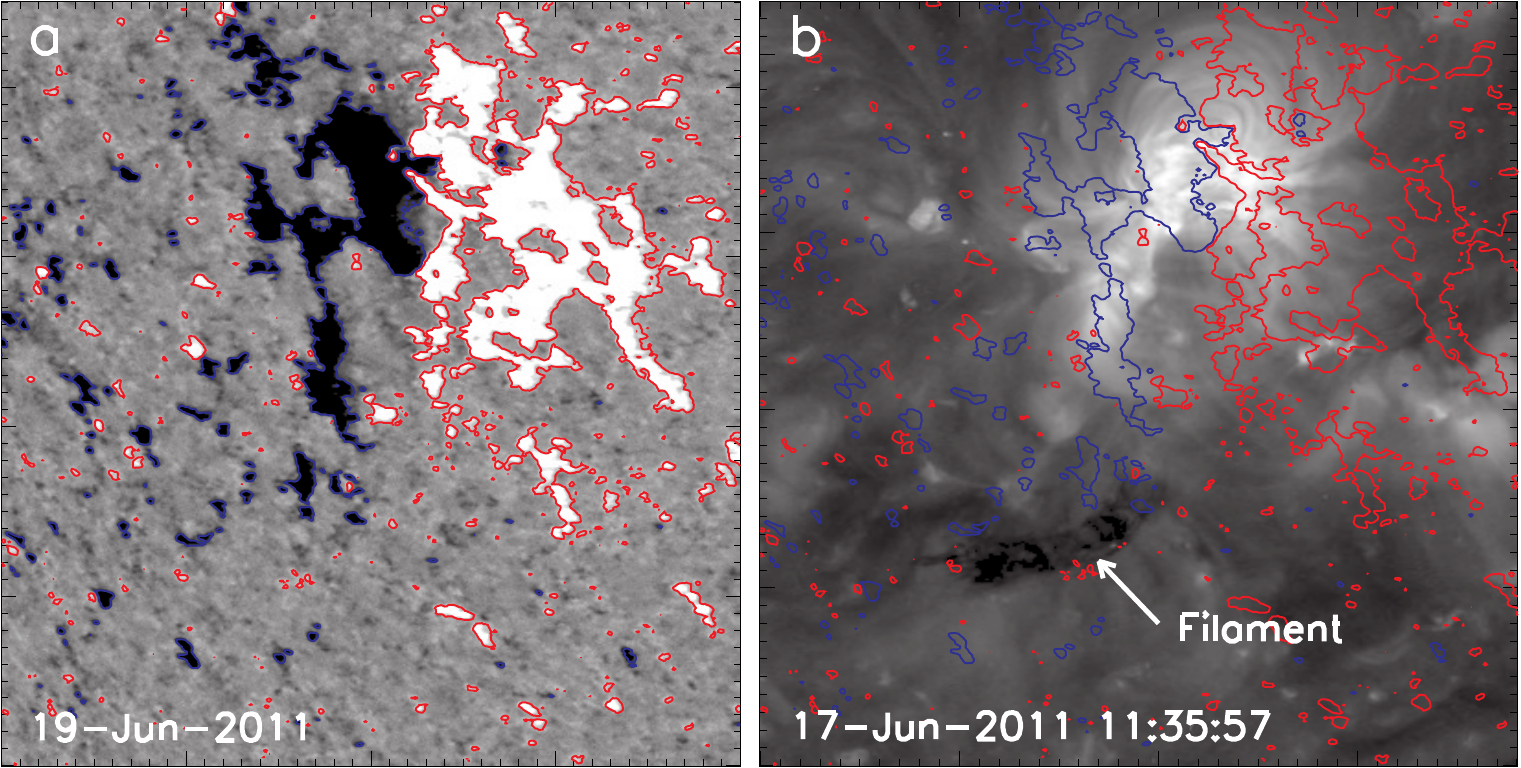}
     \caption{a) HMI magnetogram of the active region observed on 19 June 2011; b) EUVI-B 195 \AA\ image of the active    region and filament on 17 June 2011, overlaid with the contours of the HMI magnetogram shown in (a). In (a) and (b), the blue and red contours (of $\pm$ 40 G) represent negative and positive magnetic
      flux, respectively.}\label{mag}
\end{figure}

 In Figure~\ref{mosac}c, we show a light curve of a sub-region of the EUVI-B images from 16 to 17 June 2011. This is
 an integrated intensity plot of the region inside the box shown in
  Figure~\ref{4sdo}b. The intensity profile shows the apparent increase in intensity during each of the nine flares, F1-F9.
  The AIA and EUVI-B intensity time-series images
  show the flare brightenings occurring just after the flares (e.g. dashed line during  F1 in Figure~\ref{mosac}a,b) because the fiducial lines
 are not centered in the core of the active region. The EUVI-B 195\AA\ light curve also shows several other intensity peaks; these are
 due to the sudden increase in brightening in the north of the active region loops.
 Figure~\ref{mosac}c shows multiple peaks during F7. We checked this carefully and observed that there are
 two more apparent brightenings in the active region (at $\sim$ 14:00 and 15:50 UT) between F7 and F8.
 %The F7 starts at $\sim$ 13:00 UT, which quickly led to the two more brightenings in the active region at $\sim$ 14:00 and 15:50 UT.
 These brightenings are prominent in the AIA 94\AA\ images. It seems as if
 there was a cascade of loop brightenings between 13:00 and 16:00 UT, explaining the multiple peaks during F7 in Figure~\ref{mosac}c.
 We also find that there is an intensity drop around F9 (in Figure~\ref{mosac}c), which is due to the dimming before and after F9 (see Figure~\ref{dim}).
 We can also compare the integrated intensity plot of Figure~\ref{mosac}c with the
  soft X-ray flux measured by \textit{GOES} in Figure~\ref{flux}. The light curve for the EUVI-B 195\AA\ channel
 shows peak times of the nine flares to be
 at the same times as the \textit{GOES} peak times.

 \subsection{\textit{Prominence eruption}} \label{promi}

 Examining the motion of the prominence, we find that there is a slow-rise after F4, near 20:00 UT
 (Figure~\ref{mosac}a), followed by a relatively flat plateau (before F6) in the prominence slow-rise phase.
 After the sixth flare, F6, which was at $\sim$ 07:26 UT, the prominence/filament starts to rise again (see Figure~\ref{mosac}a,b).
 This slow motion continues for the next $\sim$ 6 hours.
 There is an unambiguous rise in the prominence/filament height after F7, at 13:50 UT. With F7, the prominence/filament becomes destabilized
 and begins to rise further at a relatively faster speed; we say that the prominence/filament
 enters its fast-rise phase at this time. It subsequently reaches to a maximum height, before the prominence/filament plasma drains
  down towards the solar surface.
  The fast-rise of the filament lasts for 3 hours, between $\sim$ 13:50 and 17:05 UT\@.
 Finally, it detaches from the solar surface with the last homologous flare, F8, which peaked at 17:08 UT
  (according to the RHESSI\footnote{$http://hesperia.gsfc.nasa.gov/hessidata/dbase/hessi$\_$flare$\_$list.txt$} flare list).
 We also measured the velocities of the prominence rise during its slow-rise and fast-rise
 by following the motion of the prominence along the fiducial line (in Figure~\ref{4sdo}a).
 The prominence moves with an average velocity of
 $\sim$ 1 \kms\ between 07:30 UT and 14:00 UT during its slow-rise phase. During the fast-rise phase (after F7), it moves with
 an average velocity of $\sim$ 4 \kms.

 To see the evolution of the prominence during its eruption,
 we create a distance-time image along the white diagonal line in Figure~\ref{reconnection}b.
 Figure~\ref{reconnection1}a,b shows the upward motion of the prominence, which starts at $\sim$ 13:50 UT.
 A brightening appears near the top of
 the prominence at $\sim$ 15:30 UT (see MOVIE193).
 Strong downflows appear in 171\AA\ and
 193\AA\ after the brightening at the top (see Figure~\ref{reconnection1}). We suspect that
 the brightening was due to magnetic reconnection between the field containing the flare ejecta, and the field carrying the filament material
  during F7.
 This brightening is clearly visible in AIA 304, 171, 193 and 211 \AA\ channels, but not in the 94 \AA\ channel.
 This suggests that these brightenings were low-level ($\sim$ 2 MK) heating events.
 Finally, the prominence material falls back towards the solar surface.
 Possibly, the field containing the PFCS did
 not allow the prominence material to escape.
 At least at this time, there was no complete opening of the field lines.

\subsection{\textit{Coronal dimming and the CME}} \label{dimming}

 Figure~\ref{arc}a,b shows the expansion of the overlying coronal field above the
prominence/filament system. The biggest change
in that overlying coronal field occurred after F8; F8 erupted at 17:14 UT on 17 June 2011. The overlying field lines
started to expand slowly at 17:50 UT (see MOVIE193).  They opened completely at about the time of the last flare, F9, at $\sim$
21:00 UT (see Figure~\ref{4sdo}d). At the same time, we notice a strong dimming to the south of the active region in  the EUVI-B 195\AA\ images.

To see the evolution of the
dimming with time, we have created a base-difference time-series image (see Figure~\ref{dim}) along the vertical dashed line shown in
Figure~\ref{4sdo}b. For the base image, we averaged $\sim$ 1.5 hours of data
prior to F7. The dimming can be clearly seen in the base-difference image.
 Figure~\ref{dim} shows both weak and strong dimmings. We notice an appearance of weak dimming just after F8,
which is due to expansion of coronal field lines after the filament eruption. Simultaneously, this dimming was observed in the
AIA 193 \AA\ images (shown with an arrow in Figure~\ref{mosac}a).
The strong dimming appears around 21:05 UT (in Figure~\ref{dim}).
It occurred after F9, and is clearly a consequence of that F9 eruption.

From the SOHO/LASCO C2 coronograph, a CME was observed at 00:23 UT on 18 June 2011 (Figure~\ref{arc}c). %\footnote{$http://cdaw.gsfc.nasa.gov/movie/make_javamovie.php?stime=20110617_2236&etime=20110618_0529&img1=lasc2rdf&title=20110618.003606.p119g;V=234km/s$}.
We expect that the CME originated from the strongly-dimmed region. It is known
 that dimmings often occur in conjunction with flares and filament eruptions \citep[e.g.][]{sterling97,harra01,howard04},
and dimmings are mainly due to density
depletion along the line-of-sight \citep{webb12}. From our observations, it appears that
the strong dimming results from the removal of coronal material with the last eruption, F9.

\section{Discussion and Summary} \label{Discussion}
Eight homologous eruptive flares were observed by SDO/AIA and STEREO/EUVI-B in active region NOAA 11237 over 16 - 17 June 2011.
We observed a prominence/filament
cavity system (PFCS) that was magnetically connected to the flaring region. The PFCS was continuously disrupted by
ejecta expelled into it by the eight homologous eruptions from the nearby active region.
 We mainly diagnose the 3D dynamics of the
PFCS during the series of homologous flares.
Over the two-day time period investigated, we found that with each of the first eight
homologous flares (F1-F8), some hot material, presumably together with magnetic field, was ejected
from the core of the active region. This material and field entered and interacted with the PFCS; this is
clearly seen in Figure~\ref{over} and \ref{over1}. This caused the PFCS to rise upwards slowly but
significantly in several steps (after F4, F6 and F7).

During the first three flares (F1-F3) on 16 June 2011, the prominence is not fully discernible in the AIA images.
Apparently it was beyond the limb, and we confirmed this by looking at in the STEREO images of the filament on the disk
 (Figure\ref{over}b,d,f and MOVIE304). At later times, the prominence response to the flares is clearly discernible in the AIA limb images.
By following the motion of the prominence in the AIA 193\AA\ images, we find conspicuous jumps in the prominence trajectory after
F4, F6 and F7. A slow-rise motion in the prominence trajectory is observed for $\sim$ 4-5 hours after F4 ($\sim$ 20:00 UT), followed by
a plateau in velocity for $\sim$ 4.5 hours after F5.
After F6, there is a continuous slow-rise in the prominence/filament trajectory for about 6-7 hours
  followed by a fast-rise (after F7), for 3 hours,
 as observed in the SDO and STEREO images. 
 Similar behavior of filament transition from
slow-rise to fast-rise has also been studied by e.g., \cite{sterling05} and \cite{sterling12}. In the present study, the slow-rise and
fast-rise are apparently triggered by the
flare eruptions from the active region.

\begin{figure}
   \centering
  \includegraphics[width=\linewidth]{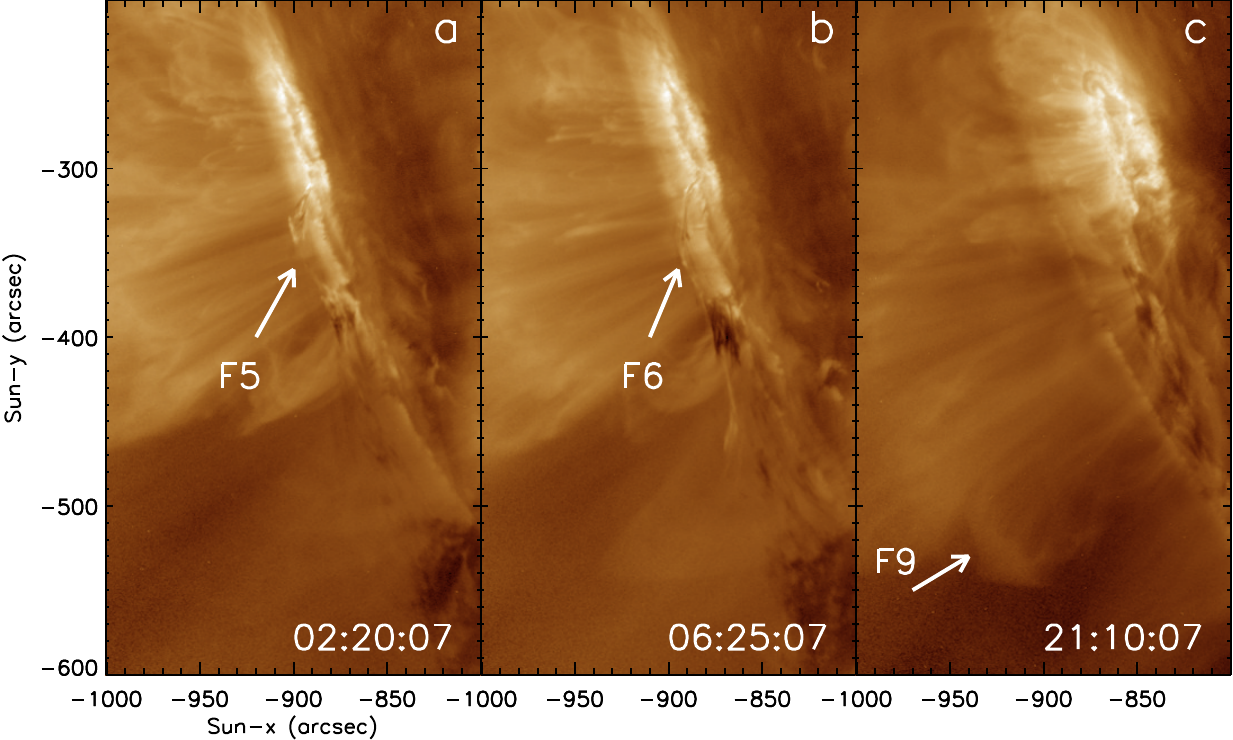}
     \caption{Flares on 17 June 2011: (a-c) AIA 193 \AA\ intensity images. In (a), (b) and (c), the arrows point to the
     F5, F6, and F9, respectively.} \label{flares}
\end{figure}

 Bright ejecta are expelled during each homologous eruption and move along the cavity field lines, and each ejection disturbed
 the magnetic field holding down the PFCS. Some faint brightenings
 are observed at the remote filament footpoint (arrow in Figure~\ref{4sdo}b) in EUVI-B 304 \AA\ and 195 \AA\ images.
 The prominence/filament begins to rise fast with F7 and it becomes detached completely
 from the solar surface after F8. Shortly after F8, prominence plasma drains to the solar surface but the overlying field
 of the PFCS continued to move outward as indicated by the weak dimming in Figures~\ref{mosac}a and \ref{dim}. %This eruption successively removed the overlying field containing the PFCS and led to a final ejective eruption, F9, followed by a CME.
  Basically, each of the eight homologous eruptions gradually weakened the magnetic tension force that maintains the overall stability of the PFCS.
 The last homologous eruption, F8, removed the enveloping field above the PFCS. Part of that enveloping
 field also covered the nearby active region, and removal of that field resulted in the final eruption from that active region, F9, which was ejective.

 Figure~\ref{flares} shows a close-up of the core portion of the active region, which was the site of the
origin of flares F1--F9.  That figure compares the situation with the confined eruptions, F1--F8,
and that of the ejective eruption F9.  Figure~\ref{flares}a,b shows the situation that was typical for
the confined eruptions:  a small-scale ($\sim$50\arcsec) filamentary feature in the core region
contained a mixture of absorbing and emitting material.  This feature was dynamic, as is apparent
from the movies (MOVIE193 and MOVIEAIA), and was located where the principle flare brightenings
occurred corresponding to each of the respective confined eruptions.  Figure~\ref{flares}a,b  shows
the behavior of flares F5 and F6, and accompanying MOVIEAIA almost shows that each of the flares 
F1-F8 display essentially same basic behavior.
In contrast, flare F9 showed different behavior in the core region of the active region (see Figure~\ref{flares}c).
This morphological difference between F1--F8 on the one hand, and F9 on the other, suggests
that either the triggering mechanism for F9 was different from that of the first eight eruptions, or the
triggering mechanism was the same but the physical consequences of that mechanism likely were different due to 
 a change in the local magnetic topology resulting from the removal of the overlying field.
 These physical consequences include the morphological
differences illustrated in Figure~\ref{flares}, and another is the production of a CME.
During F9, ejected material escaped the Sun as a CME, shown with an
 arrow in Figure~ \ref{arc}c and \ref{flares}c. In contrast, during the first eight
 eruptions (F1-F8), the ejected material was trapped in the overlying coronal field,
 which did not allow the material to leave the Sun (Figure~\ref{flares}a,b).
 We speculate that the eruption F8 successively removed the overlying field, triggering a final ejective eruption F9,
 followed by a CME. It is worth mentioning here that last two flares (F8 and F9) might be called sympathetic-like flares
 because F8 causes the filament to erupt, and then that eruption leads to last flare F9.

\begin{figure*}
   \centering
  \includegraphics[width=\linewidth]{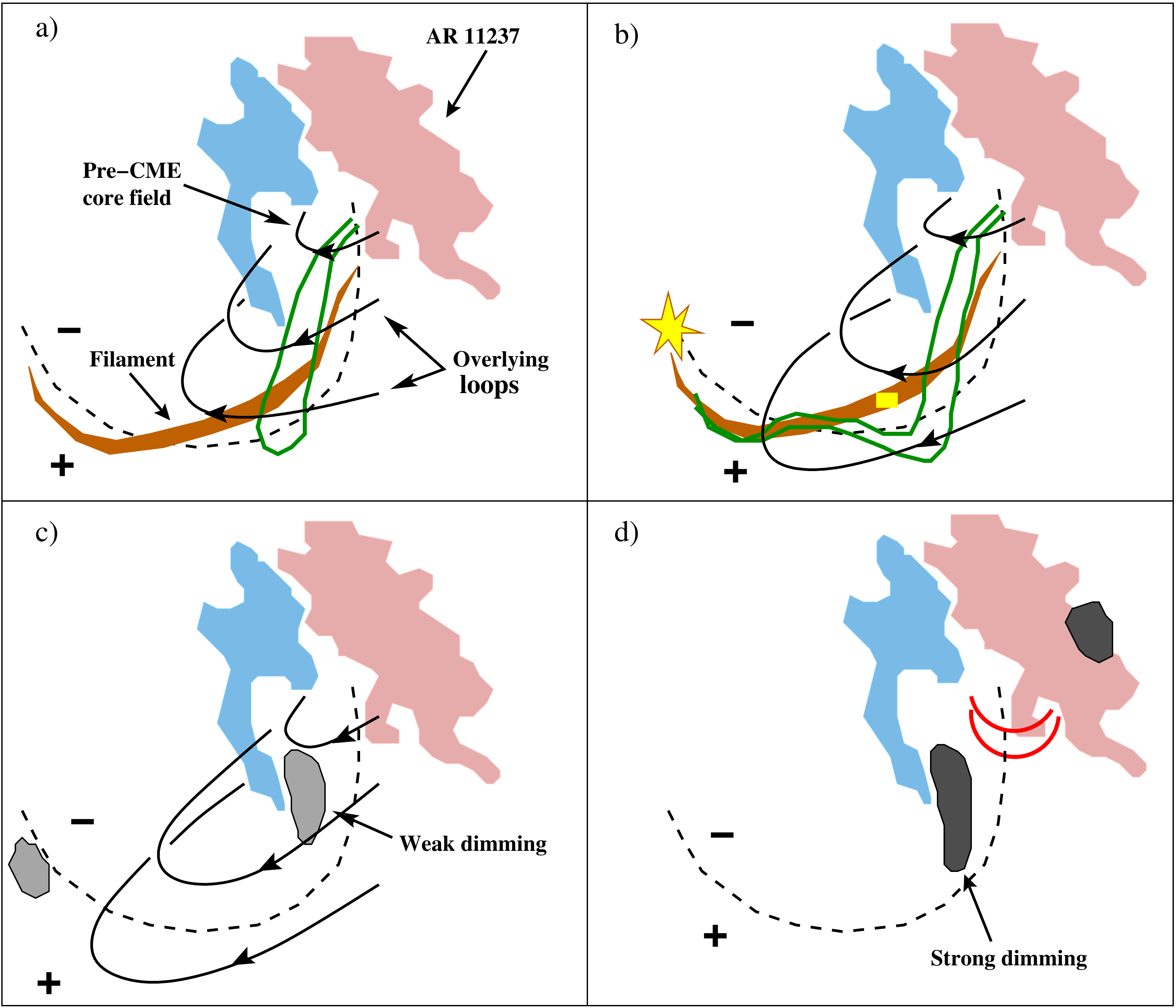}
     \caption{Schematic illustration of observations: A STEREO/EUVI-B view of the filament (orange) and active region.
     Blue and pink patches indicate
     negative and positive magnetic flux, respectively. The black dashed line roughly outlines the position of the photospheric neutral line.
     In (a) and (b), the green loop represents material and field ejected from the core of the AR during the eight homologous-flare
     eruptions, resulting in the reconnection (yellow box in b) with the field holding the filament
     (the PFCS) and also resulting in an expansion and eventual
      opening of overlying coronal loops in (b) and (c), followed by a weak coronal dimming in (c). Panel (d) shows the
      situation following complete removal of overlying loops and the
     strong coronal dimming after the last eruption, F9; the red arcs indicate the F9 flare loops}\label{skt}
 \end{figure*}

Flare F9 follows the catastrophic disruption to the PFCS by flare F8.  Unlike the first eight eruptions
(F1-F8), F9 was {\it not} confined.  Rather, it apparently was an ejective eruption, resulting in
the CME (Figure~\ref{arc}c), and in strong coronal dimming (Figure~\ref{dim}). F9 appears to have originated from
approximately the same location in the core of the active region as F1-F8.
 So why were F1-F8 confined and F9 ejective?  We speculate that, when the
PFCS was disrupted after F8, the field enveloping the PFCS started to erupt slowly; that
is, the field slowly (in 3.5 hours) began to open (see MOVIE193 at 18:55 UT).
This slow-field-opening process removed some of the field that had
been overlying and confining the F1-F8 flares.  With this overlying field removed, the next flare eruption,
F9, was able to escape the Sun without being confined. This process is similar to what has been observed by 
\cite{schrijver11} and modeled by the \cite{torok05}. The process where by overlying field is removed by the first eruption and results in a
second eruption was called `lid removal' in \cite{sterling14}.
In the present study, the so-called `overlying lid/field' had been removed by the
last homologous eruption, F8, which allowed the F9 eruption to start and to not be confined.

 In Figure~\ref{skt}, we summarize our observations with the help of a cartoon.
 It shows the evolution of the event according to the SDO and STEREO observations. We do not have magnetic field information
 during our observation period and
 so the details of this summary are inevitably speculative.
 Figure~\ref{skt}a represents the situation prior to the final three eruptions (F7 to F9). It shows the filament lying along
 the photospheric neutral line with overlying
 coronal field lines. A pre-CME core field is shown with an arrow in Figure~\ref{skt}a and Figure~\ref{flares}c.
 The green field shows the path taken by hot material that ejected
 from the active region core during the homologous eruptions.
 Each ejection hit the filament field and destabilized it.
 More precisely, with each homologous eruption, the flare ejecta, which consisted  of hot material
and magnetic field in which that material was entrained, struck in a broadside fashion the PFCS field
that contains the filament. This concept of the ejecta consisting of hot material 
and magnetic field is fully consistent with the standard model for solar eruptions. 
According to the standard model, an eruption expels filament material and field, and both eventually becomes part of a CME 
(see Figure 1 of \cite{moore01} and Figure 1 of \cite{shibata95}). In the case of our homologous eruptions, we are proposing that 
our `ejecta' correspond to the CME-producing components of this 
standard-model:  the `hot material' of the ejecta corresponds to the expelled filament in the 
standard model (we are assuming that the filament material underwent some heating during the eruption), and the `field' of the ejecta 
corresponds to the field surrounding the expelled filament in the standard model. In the standard model as depicted in the above references,
 the expelled filament and field move out vertically from the Sun. In our homologous flares, the ejecta instead moves out nearly horizontally, 
 and becomes trapped on the larger scale field that wraps around the PFCS.
 
 Figure~\ref{skt}b,c,d shows the filament system after F7, F8, and F9 respectively.
 The coronal cavity started to expand slowly
 with the eruption F7, which allowed the filament material to rise further. At the same time, there was
 a brightening at the remote filament footpoint (in weak magnetic field), shown by the star in Figure~\ref{skt}b. %It has been noticed in many observations filament undergo a period of `brightening' before the eruption  \citep[e.g.][] {harrison03}.
 During each eruption there was an ejection of hot material, as discussed before, which moved along the cavity field
 and brightened the remote footpoint of the filament (see Figure
 \ref{skt}b and accompanying MOVIE193). The filament slowly started moving away from its remote footpoint
 just after the brightening (after F7 at 14:35 UT) and left a dimming (Figure~\ref{skt}c).
 In addition to footpoint brightening,
  one more enhanced brightening is seen at the top of the prominence at 15:30 UT (approximately at the location of the yellow box
  in Figure~\ref{skt}b; corresponding to the location indicated by the arrow in Figure~\ref{reconnection}b). The prominence material
  below the brightening is observed draining towards the surface. Plausibly, the brightening was due to the spatially localized magnetic reconnection
  between the field containing the flare ejecta, and the field carrying the filament material.
  This reconnection possibly
  reconfigured the field lines and led to the observed downflow. The overlying coronal field did not allow the prominence/filament
  material to escape outward and led to a prominence/filament eruption \citep{moore01}. After $\sim$ 17:50 UT, the overlying field lines
  (labeled as `overlying loops' in Figure~\ref{skt}a) visibly expanded over the
  next 3.5 hours (shown in Figure~\ref{skt}b,c), leading to the weak dimming, and to the last flare, F9,
  at 21:00 UT (flare loops are shown in Figure~\ref{skt}d, and Figure~\ref{4sdo}d; they are also visible in MOVIE193).
  This suggests that the removal of the field overlying the active region may have triggered the final `ejective' eruption from the core of
  the active region.
  Shortly after F9, a region of strong dimming (strong compared to the earlier weaker dimming) occurred close to
  the negative and positive flux of the active region (Figure~\ref{skt}d). The appearance of the strong dimming over a large surface area
  just after the flare was presumably due to the expulsion of coronal
  material with the CME \citep{harrison00,howard04}. The CME was seen at 00:37 UT on 18 June 2011, which was after the
   strong dimming onset and 3 hours after the start of flare F9.
   We speculate that the series of homologous eruptions
    destabilized the PFCS and thereby opened the overlying coronal field, leading to the CME.

\acknowledgments
%\begin{acknowledgements}
%\center
%\acknowledgments {\bf Acknowledgements}\\
We acknowledge the use of the SDO/AIA,
STEREO/EUVI observations for this study.
SDO data are courtesy of the NASA/SDO AIA
and HMI science teams. STEREO data are courtesy
of the STEREO Sun Earth Connection Coronal and Heliospheric Investigation (SECCHI)
team. This work was supported by the Heliophysics Division
of NASA's Science Mission Directorate through the Living With a Star Targeted Research and
Technology Program, and by the Hinode Project.
%\end{acknowledgements}
%\email{aastex-help@aas.org}.

%%

\bibliographystyle{apj}

\end{document}